\documentclass[aps,pra,twocolumn,superscriptaddress,showpacs]{revtex4-1}

\usepackage{amsfonts}
\usepackage{amsmath}
\usepackage{amssymb}
\usepackage[pdftex]{graphicx}
\usepackage{bm}

\begin{document}

\title{Crossover in the magnetic response of single-crystalline Ba$_{1-x}$K$_x$Fe$_2$As$_2$ and Lifshitz critical point evidenced by Hall effect measurements}

\author{Yong Liu}
\email[Corresponding author: ]{yliu@ameslab.gov}
\affiliation{Division of Materials Sciences and Engineering, Ames Laboratory, Ames, Iowa 50011, USA}

\author{Thomas A. Lograsso}
\affiliation{Division of Materials Sciences and Engineering, Ames Laboratory, Ames, Iowa 50011, USA}
\affiliation{Department of Materials Science and Engineering, Iowa State University, Ames, Iowa 50011, USA}

\date{\today}

\begin{abstract}
We report the doping evolution of magnetic susceptibility $\chi (T)$ and Hall coefficient $R_H$ in high quality Ba$_{1-x}$K$_x$Fe$_2$As$_2$ ($0.13 \leq x \leq 1$) single crystals. It is found that the normal-state magnetic susceptibility of Ba$_{1-x}$K$_x$Fe$_2$As$_2$ compounds undergoes a crossover from linear-$T$ dependence in the undoped and underdoped samples into KFe$_2$As$_2$-type magnetic response in the overdoped samples with increasing K content. Although magnetic susceptibility $\chi (T)$ of optimally doped samples ($0.34 \leq x \leq 0.47$) still follows a monotonic increase with increasing temperature, a big hump around 300 K emerges. As $x$ exceeds 0.53, a broad peak forms in overdoped samples ($0.53 \leq x \leq 1$), which shifts toward 120 K for the end member KFe$_2$As$_2$. Above the peak temperature $T$*=120 K, a Curie-Weiss-like behavior is observed in KFe$_2$As$_2$. Hall coefficient $R_H$ of underdoped sample $x$=0.22 shows a rapid increase above spin-density-wave transition temperature $T_{SDW}$. Below $T_{SDW}$, it increases slowly. $R_H$ of optimally doped and slightly overdoped samples ($0.34 \leq x \leq 0.65$) shows relative weak temperature dependence and saturation tendency below 150 K. However, $R_H$ of K heavily overdoped samples ($0.80 \leq x \leq 1$) increases rapidly below 150 K. Meanwhile, Hall angle $cot \theta _H$ displays a concave temperature dependence within the doping range $0.22 \leq x \leq 0.55$, whereas it changes to a convex temperature dependence within the doping range $0.65 \leq x \leq 1$. The dramatic change coincides with Lifshitz transition occurred around the critical doping $x$=0.80, where angle photoemission spectroscopy measurements had confirmed that electron pocket disappears with excess hole doping in Ba$_{1-x}$K$_x$Fe$_2$As$_2$ system. It is suggested that the characteristic temperature $T$* at around $120 \sim 150$ K observed in susceptibility and Hall coefficient, as well as previously reported resistivity data, may indicate an incoherence-coherence crossover in Ba$_{1-x}$K$_x$Fe$_2$As$_2$ system.
\end{abstract}


\maketitle

\section{Introduction}

Novel magnetism and multiband structure are two key aspects in the research of iron-based superconductors \cite{Johnston,Stewart,Lumsden,PDai,Hirschfeld,Chubukov}. Parent compounds such as LaOFeAs and BaFe$_2$As$_2$ show a spin-density-wave (SDW) transition at $T_{SDW}$~140 K \cite{JDong,Rotter-BFA}, coupled with a phase transition from tetragonal to orthorhombic structures. The normal state of iron-based superconductors is a strongly correlated metal and the parent compound is a bad metal at the verge of the metal insulator transition \cite{Haule08}. By aliovalent and isovalent ions doping or an application of pressure, the SDW order is suppressed, while a superconducting dome emerges with increasing doping levels in the phase diagram \cite{Johnston,Stewart}. The primary pairing interaction was proposed to be mediated by antiferromagnetic (AFM) spin fluctuations. As a result, the superconducting state was expected to be $s_\pm$ state, i.e., extended $s$-wave pairing with a sign reversal of the order parameter between different FS sheets \cite{Mazin}. Among the iron-based superconductors, Ba$_{1-x}$K$_x$Fe$_2$As$_2$ system is quite unique. The optimally doped sample $x$=0.4 displays a $T_c$ of 38 K. With increasing K doping level, $T_c$ steadily decreases to 3.8 K for the end member KFe$_2$As$_2$ \cite{Rotter-phase-diagram}. It was found that the electronic structure of Ba$_{1-x}$K$_x$Fe$_2$As$_2$ compounds shows a dramatic change from optimally doped to overdoped samples \cite{Malaeb,Sato}. Accompanied with the evolution of electronic structure, the pairing symmetry seems to change from $s_\pm$ wave in optimally doped samples to $d$ wave in KFe$_2$As$_2$ \cite{Reid}. Recent Angle resolved photoemission spectroscopy (ARPES) found that Fermi surface (FS) topology of Ba$_{0.1}$K$_{0.9}$Fe$_2$As$_2$ single crystal is similar to that of KFe$_2$As$_2$, but differs from that of Ba$_{0.3}$K$_{0.7}$Fe$_2$As$_2$, which was interpreted within the framework of Lifshitz transition occured between $0.7<x<0.9$ \cite{Xu}. Theoretical calculations also pointed out that dissolution of electron cylinders occurs near $x \sim 0.9$ with Lifshitz transition in Ba$_{1-x}$K$_x$Fe$_2$As$_2$ superconductors \cite{Khan}. The doping dependent FS reconstruction is also evidenced by the change of thermoelectric power $S_{ab}$ for overdoped Ba$_{1-x}$K$_x$Fe$_2$As$_2$ single crystals, where the maximum at around 120 K in temperature dependence of $S_{ab}$ collapses into a plateau at $x\sim 0.8-0.9$ \cite{Hodovanets-TEP}.

The transport property of Ba$_{1-x}$K$_x$Fe$_2$As$_2$ system also shows different behavior, compared to electron doped Ba(Fe$_{1-x}$Co$_x$)$_2$As$_2$ and isovalent doped BaFe$_2$(As$_{1-x}$P$_x$)$_2$. A linear-$T$ dependence of in-plane resistivity $\rho_{ab}$ was universally observed in the optimally doped BaFe$_2$(As$_{1-x}$P$_x$)$_2$ \cite{Kasahara}, Ba(Fe$_{1-x}$Co$_x$)$_2$As$_2$ \cite{Tam}, and Ba(Fe$_{1-x}$Ni$_x$)$_2$As$_2$ \cite{Zhou} single crystals, while the Fermi liquid behavior $n \sim 2$ was observed in overdoped regime by a fit of power law $\rho_{ab}=\rho_{0}+AT^{n}$. It is noted that the exponent $n \sim 1.5$ in optimally doped Ba(Fe$_{1-x}$Ni$_x$)$_2$As$_2$ samples was reported by different group \cite{HLuo}. For Ba$_{1-x}$K$_x$Fe$_2$As$_2$ single crystals, however, it was found that $\rho _{ab}$ actually follow $T^{1.5}$ dependence in the optimally doped regime. And $T^2$ term contributes a lot in the entire doping range $0.22 \leq x \leq 1$ \cite{Liu-BKFA}. In an early report on the transport properties of Ba$_{1-x}$K$_x$Fe$_2$As$_2$ single crystals within low K doping regime ($0 \leq x \leq 0.4$), it was found that the power exponent $n$ evolves from 2 for the undoped samples to 1 at optimal doping $x$=0.37 \cite{Shen}. The discrepancy may result from different temperature windows for the fits of power law and quality of single crystals. Furthermore, all superconducting Ba$_{1-x}$K$_x$Fe$_2$As$_2$ samples from underdoped to overdoped regimes show a saturation tendency above 100 K \cite{Liu-BKFA}.

In this study, we report the doping evolution of normal-state magnetic susceptibility, and Hall coefficient, and Hall angle in Ba$_{1-x}$K$_x$Fe$_2$As$_2$ ($0.13 \leq x \leq 1$) single crystals. We find that magnetic susceptibility $\chi(T)$ monotonically increases with increasing temperature for the underdoped and optimally doped samples $0.13 \leq x \leq 0.47$. A broad peak emerges as $x$ exceeds 0.53, which suggests different magnetic interactions in the overdoped regime. Intriguingly, we observed a dramatic change of Hall coefficient $R_H$ and Hall angle $cot\theta_H$ as $x$ crosses the doping $x$=0.80, where Lifshitz transition occurs with the change of FS topology evidenced by ARPES measurement \cite{Xu} and suggested by theoretical calculations \cite{Khan}.

\section{Experimental details}
High quality Ba$_{1-x}$K$_x$Fe$_2$As$_2$ ($0.13 \leq x \leq 1$) single crystals were grown by using self-flux method \cite{Liu-BKFA,Liu-KFA}. The crystals can be easily cleaved into thin plates along $ab$ plane. Magnetic susceptibility $\chi (T)$ and Hall resistivity $\rho_{xy}$ were measured by using Physical Property Measurement System (PPMS, Quantum Design). For the measurements of magnetic susceptibility, the magnetic field $H$ was applied parallel to the $ab$ plane ($H \parallel ab$) and perpendicular to the $ab$ plane ($H \parallel c$). Nearly ten pieces of crystals with amount of $20\sim 40$ mg were piled along $c$ axis for each measurement. In order to further clarify the intrinsic magnetic response of the samples, the magnetization as a function of applied field $H$ was measured at a series of fixed temperatures. The temperature dependence of magnetic susceptibility curves was verified by the susceptibility data extracted from field dependent behavior. For the high-temperature susceptibility measurements, the crystals were glued on a heat stick (PPMS VSM oven) by using cement.

The Hall resistivity $\rho _{xy}$ was measured in magnetic field dependence at fixed temperatures. Because of the small Hall signal, misaligned contacts lead to a significant contribution to Hall voltage from the longitudinal resistivity $\rho _{xx}$. In order to avoid this problem, the Hall signal can be extracted from the slope of linear filed dependence of Hall voltage by sweeping magnetic field. The Hall coefficient is then calculated as $ R_{H}=\frac{V _{H} \times d}{I _{s} \times H} $, where $V _{H}$ is Hall voltage, $d$ is thickness of the thin plate-like crystals, $I _{s}$ is driven current, and $H$ is applied magnetic field. The thin flakes with a thickness of 10-30 $\mu$m were obtained by peeling off single crystals using adhesive tape. Five probe contacts were made by soldering the gold wires to the single crystals. The driven current of 1 mA and 19 Hz was used in the Hall effect measurements. Two pieces of crystals were measured for each K doping to check the reproducibility of the Hall data.

\section{Results and discussion}

\begin{figure}[tbh]%
\centering
\includegraphics[width=8cm]{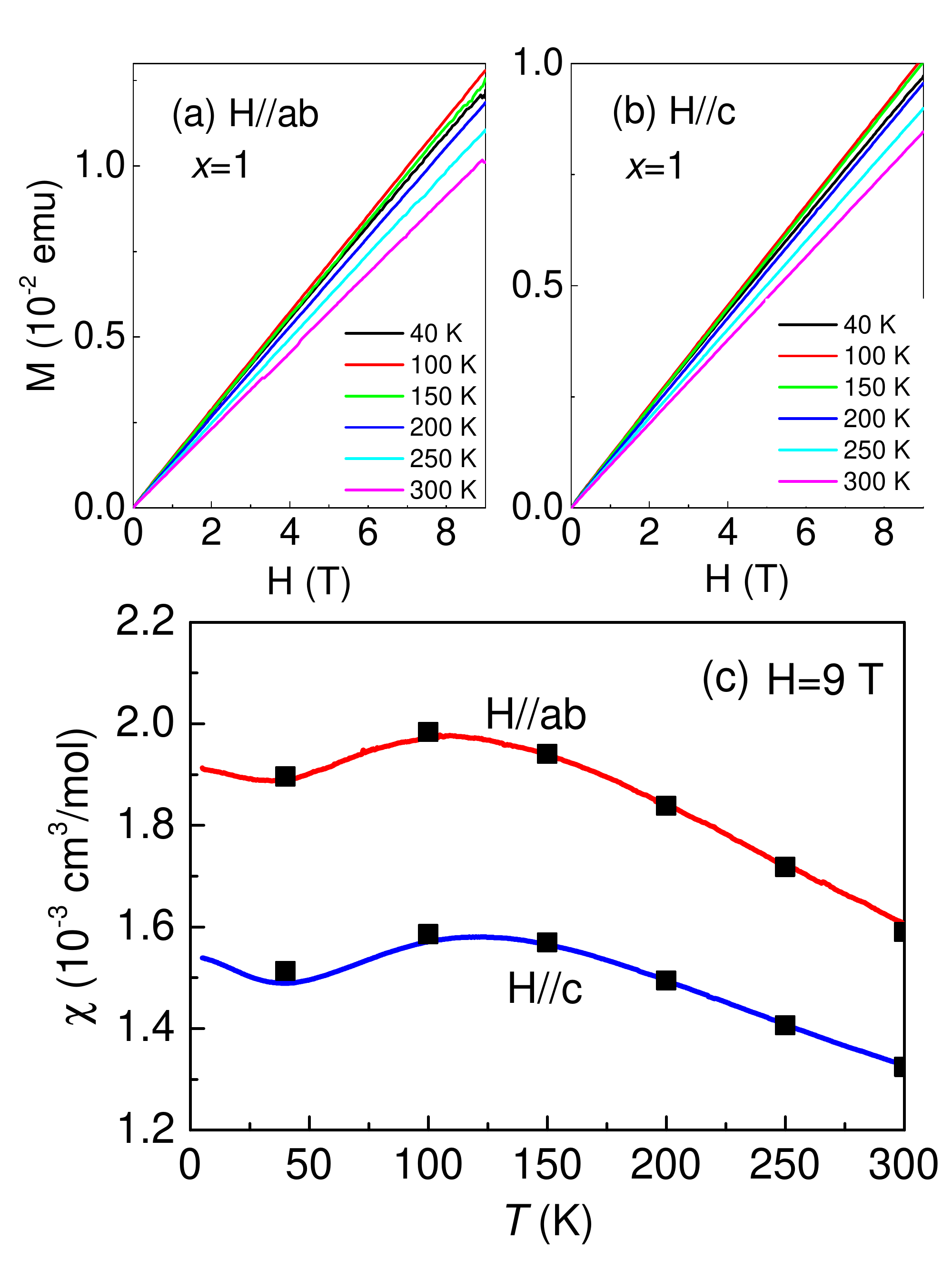}%
\caption{(Color online) Isothermal magnetization curves of KFe$_2$As$_2$ single crystal for (a) $H \parallel ab$ and (b) $H \parallel c$, measured at 45, 100, 150, 200, 250, and 300 K. (c) Temperature dependence of magnetic susceptibility $\chi(T)$ of KFe$_2$As$_2$ single crystal is measured by an application of magnetic field of 9 Tesla, represented by solid lines. Solid squares correspond to the susceptibility data obtained from the linear fit of isothermal magnetization curves.}%
\label{MKFA}%
\end{figure}

It is important to clarify the intrinsic magnetic response of iron-based superconductors because they may contain ferromagnetic inclusions \cite{Johnston}. Figures \ref{MKFA}(a) and \ref{MKFA}(b) shows the isothermal magnetization curves of KFe$_2$As$_2$ single crystal measured at 45, 100, 150, 200, 250, and 300 K in the configurations of $H \parallel ab$ and $H \parallel c$. A linear field dependence of magnetization $M$ rules out the existence of magnetic impurity phases. Magnetic susceptibility $\chi$ is defined as $\chi = \partial M/\partial H$, i.e. the slope of the $M$ vs $H$ curves. Temperature dependence of the magnetic susceptibility $\chi (T)$ of the same sample was measured under a magnetic field of 9 T, as shown in Fig. \ref{MKFA}(c). As can be seen, the susceptibility data extracted from the linear fit of isothermal magnetization curves fall on the temperature dependent curve. A broad peak emerges at around 120 K, which is consistent with the previous results by Hardy {\it et al} \cite{Hardy}.

\begin{figure}[tbh]%
\centering
\includegraphics[width=8cm]{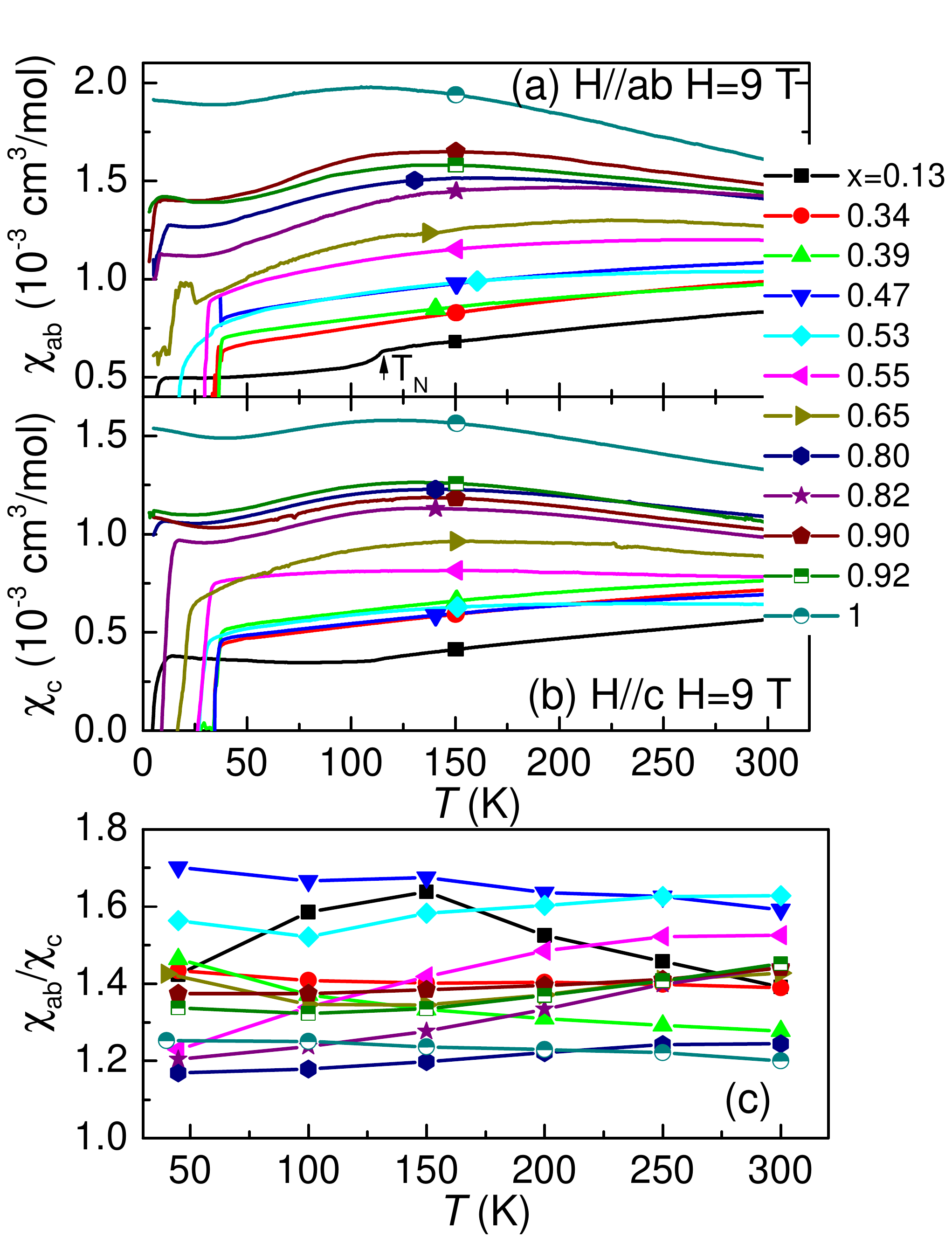}%
\caption{(Color online) Temperature dependence of magnetic susceptibility $\chi (T)$ for Ba$_{1-x}$K$_x$Fe$_2$As$_2$ ($0.13 \leq x \leq 1$) single crystals for (a) $H \parallel ab$ and (b) $H \parallel c$ measured under 9 Tesla. (c) Temperature dependence of anisotropy ratio $\chi_{ab} /\chi_{c}$ of Ba$_{1-x}$K$_x$Fe$_2$As$_2$ single crystals.}%
\label{MBKFA}%
\end{figure}

Figures \ref{MBKFA}(a) and \ref{MBKFA}(b) show temperature dependence of magnetic susceptibility $\chi (T)$ of Ba$_{1-x}$K$_x$Fe$_2$As$_2$ ($0.13 \leq x \leq 1$) single crystals for $H \parallel ab$ and $H \parallel c$, respectively. Underdoped sample $x$=0.13 displays a kink at $T_{SDW} \sim 110$ K, which matches the SDW transition temperature in the phase diagram \cite{Rotter-phase-diagram,Avci}. Above $T_{SDW}$, a linear-$T$ susceptibility $\chi (T)$ is observed. For the optimally doped samples $x$=0.34, 0.39, and 0.47, $\chi (T)$ still keep monotonic increase with increasing temperature. But the susceptibility curves display a slightly down bending behavior, not strictly following the linear relationship. With further increasing K doping levels, $\chi (T)$ curves of overdoped samples ($0.53 \leq x \leq 0.65$) flatten out, compared to a gradual fall observed in underdoped and optimally doped samples. A big hump ranging from $T_c$ to room temperature is observed. This big hump further evolves into a broad peak centered at 120 K for KFe$_2$As$_2$. A Curie-Weiss tail is observed at low temperature regime above $T_c$ for K heavily doped samples ($0.80 \leq x \leq 1$). It is noted that the magnitude of $\chi (T)$ increases from underdoped to overdoped samples, showing the similar doping dependent behavior to that observed in polycrystalline samples \cite{Storey}. In Fig. \ref{MBKFA}(c) we show the temperature dependence of anisotropy ratio of $\chi_{ab} /\chi_{c}$ for all studied crystals. As can be seen, the anisotropy ratios $\chi_{ab} /\chi_{c}$ fluctuate between 1.2 and 1.6.

\begin{figure}[tbh]%
\centering
\includegraphics[width=8cm]{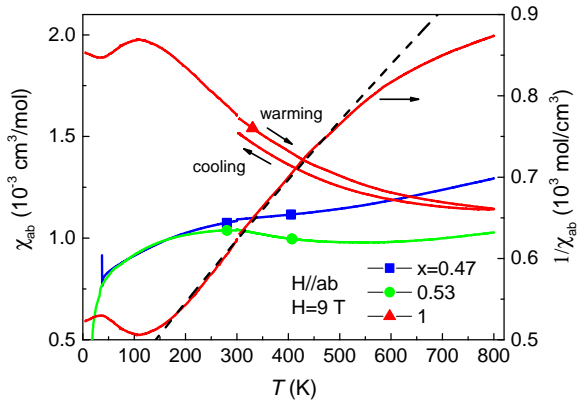}%
\caption{(Color online) High-temperature magnetic susceptibility $\chi (T)$ up to 800 K of Ba$_{1-x}$K$_x$Fe$_2$As$_2$ ($x$=0.47, 0.53, and 1) single crystals. The susceptibility data of KFe$_2$As$_2$ single crystal measured upon cooling does not follow warming curve, which should be caused by the sample degradation at high temperature. Inverse magnetic susceptibility of KFe$_2$As$_2$ single crystal is linked to the right axis. Dashed line corresponds to the fit of Curie-Weiss law. The discrepancy between the susceptibility data and Curie-Weiss law above 500 K can be explained that the sample degrades above this temperature.}%
\label{HTM}%
\end{figure}

In Fig. \ref{HTM} we show the susceptibility data measured up to 800 K for the samples $x$=0.47, 0.53, and 1. In order to identify the possible sample degradation at high temperatures, each measurement has been done on both warming and cooling processes. We find that magnetic susceptibility curves measured upon warming and cooling don not overlap each other but they still keep the similar temperature dependence, as shown in the case of KFe$_2$As$_2$. Here, we discuss the susceptibility data collected on warming process. As can be seen, the susceptibility data of the sample $x$=0.47 still follow monotonic increase with increasing temperature. Upon warming, a down bending behavior is observed, and $\chi (T)$ shows a weak hump centered at 300 K. The optimally doped samples $x$=0.34 and 0.39 shows the similar behavior (not shown in the figure). A clear broad hump is observed at around 300 K for the sample $x$=0.53, while $\chi (T)$ increases again above 550 K. For KFe$_2$As$_2$ sample, $\chi (T)$ display a broad peak at $T$=120 K. Above $T$=120 K, a Curie-Weiss-like susceptibility is observed in the paramagnetic (PM) state. The inverse susceptibility of KFe$_2$As$_2$ single crystal is shown in Fig. \ref{HTM}. The $\chi (T)$ data between $200<T<500$ K can be described by Curie-Weiss law $ \chi =\frac{C}{T - \theta_{p}} + \chi_{0}$, where magnetic parameters $C$, $\theta_{p}$, and $\chi_{0}$ correspond to the Curie constant, the Curie-Weiss temperature, and the temperature-independent contribution. The large Curie-Weiss temperature $\theta_{p}$ of -426 K suggests dominant AFM interactions for KFe$_2$As$_2$. The effective magnetic moment $\mu_{eff} \sim 2.9\mu_{B}$ was calculated from the Curie constant $C = N \mu_{eff}^2 / 3k_{B}$ ($C$=1.03). And $\chi_{0} =1.9 \times 10^{-4} $ $\textrm{cm}^3$/mol. By fixing $\chi_{0}$=0, the fit of Curie-Weiss law yields $\theta_{p}=-510$ K and $C$=1.3. The effective magnetic moment $\mu_{eff}$ is estimated to be $3.2\mu_{B}$. Hardy {\it et al.} \cite{Hardy} had reported that $\mu_{eff} \sim 2.5\mu_{B}$ and $\theta_{p} \sim -600$ K by fitting the data between $150<T<300$ K.

It is still under debate on the role of local moment in iron-based superconductors. The local Fe spin moment of parent and optimally doped CeO$_{1-x}$F$_x$FeAs ($x$=0, 0.11) and Sr(Fe$_{1-x}$Co$_x$)$_2$As$_2$ ($x$=0, 0.10) has been analysed using the Fe 3$s$ core level photoemission spectra \cite{Vilmercati}. The rapid time scales of the photoemission process allowed the detection of large local spin moments fluctuating on a 10$^{-15}$ s time scale in the PM, AFM, and superconducting phases, indicative of the occurrence of ubiquitous strong Hund’s magnetic correlations. An effective local spin $S_{eff}$ was suggested be resulted from a dynamical mixing of quasidegenerate spin states of Fe$^{2+}$ ion by intersite electron hoppings \cite{Chaloupka}. It was found that singlet correlations among $S_{eff}$ lead to the increase of the spin susceptibility with temperature. The theory can well explain the puzzle of large but fluctuating Fe moments \cite{Chaloupka}.

In Figs. \ref{MBKFA} and \ref{HTM} we already demonstrate a crossover from the linear increase to the broad peak in $\chi (T)$ of Ba$_{1-x}$K$_x$Fe$_2$As$_2$ single crystals. We notice that Co doping leads to a decrease of magnetic susceptibility of Ba(Fe$_{1-x}$Co$_x$)$_2$As$_2$ with increasing Co doping levels \cite{Wang-Co122}. In Ba$_{1-x}$K$_x$Fe$_2$As$_2$ system, however, magnetic susceptibility is enhanced with increasing K doping levels. There are already several reports on the origin of linear-$T$ dependence of $\chi (T)$ in iron base superconductors \cite{GMZhang,Korshunov,Skornyakov11,Skornyakov12}. It was suggested that strong AFM fluctuations with local SDW correlation give rise to the anomalous linear-$T$ dependence of $\chi(T)$ \cite{GMZhang}. Soon it was argued that the linear in $T$ term appears to be due to the nonanalytic temperature dependence of $\chi(T)$ in a two-dimensional Fermi liquid, which favors the itinerant scenario for iron pnictides \cite{Korshunov}. Skornyakov {\it et al.} \cite{Skornyakov11,Skornyakov12} further demonstrated linear-$T$ dependence of $\chi(T)$ in iron pnictides can be reproduced without invoking AFM fluctuations by employing the local density approximation $+$ dynamical mean field method. Furthermore, contributions to the temperature dependence of the uniform susceptibility are strongly orbitally dependent. For high temperatures ($>$1000 K) susceptibility first saturates and then decreases with temperature \cite{Skornyakov11,Skornyakov12}. Through $^{75}$As nuclear magnetic resonance (NMR) measurements on overdoped Ba$_{1-x}$K$_x$Fe$_2$As$_2$ ($x$=0.7 and 1.0) single crystals, it was found that the spin-lattice relaxation $1/T_1$ dramatically increases from the sample $x$=0.7 to the $x$=1.0, suggesting that another type of spin fluctuations develops as the doping close to $x$=1.0 \cite{SWZhang}. Hirano {\it et al.} \cite{HiranoJPSJ} performed $^{75}$As NMR and nuclear quadrupole resonance (NQR) measurements on Ba$_{1-x}$K$_x$Fe$_2$As$_2$ ($0.27 \leq x \leq 1$) single crystals. In the normal state, $1/T_1$ has a strong temperature dependence, which indicates the existence of large AFM spin fluctuations for all the studied crystals. Hardy {\it et al.} suggested that KFe$_2$As$_2$ is a strongly correlated material with highly renormalized values of both the Sommerfeld coefficient and the Pauli susceptibility \cite{Hardy}. The magnetic susceptibility of KFe$_2$As$_2$ can be comparable to that of the heavy fermion CeRu$_2$Si$_2$ which is PM state but close to AFM instability \cite{Hardy}. Therefore, the enhanced magnetism with increasing K content is closely related to the anomalous magnetic interactions in KFe$_2$As$_2$.

An explanation on the origin of the maximum in $\chi (T)$ of KFe$_2$As$_2$ single crystal comes from its heavy fermion feature. The large Sommerfeld constant $\gamma _n = 94 \sim 107$ mJ/mol K$^2$ reported in high quality KFe$_2$As$_2$ single crystals \cite{Abdel-Hafiez,Sergey,Hardy} implies a close relationship with heavy fermion compounds. Given that local moments exist in KFe$_2$As$_2$, the low-temperature maximum of $\chi (T)$ can be interpreted within the framework two-fluid behavior suggested for magnetic response of heavy electron materials \cite{Yang,Shirer}. The susceptibility in in heavy electron materials is suggested to be the sum of three contributions: conduction electron spins $\chi _{cc}$, local moment spins $\chi _{ff}$, and the hybridization of conduction and localized electrons $\chi _{cf}$. At high temperatures $\chi _{cc}$ is given by the temperature-independent Pauli susceptibility of the conduction electrons, and $\chi _{ff}$ is given by the Curie-Weiss susceptibility of the local moments. The heavy electron Kondo liquid emerges below the characteristic temperature, $T$*, as a collective hybridization-induced instability of the spin liquid that describes the lattice of local moments coupled to background conduction electrons. Above $T$*, $\chi _{ff}$ dominates. Below $T$*, $\chi _{cf}$ becomes significant. $T$* is determined by the effective Ruderman–Kittel–Kasuya–Yosida (RKKY) interaction between the nearest-neighbor local moments \cite{Yang,Shirer}. It is therefore suggested that the maximum in $\chi (T)$ of KFe$_2$As$_2$ indicate the growth of hybridization of conduction and localized electrons with decreasing temperature.

\begin{figure}[tbh]%
\centering
\includegraphics[width=8cm]{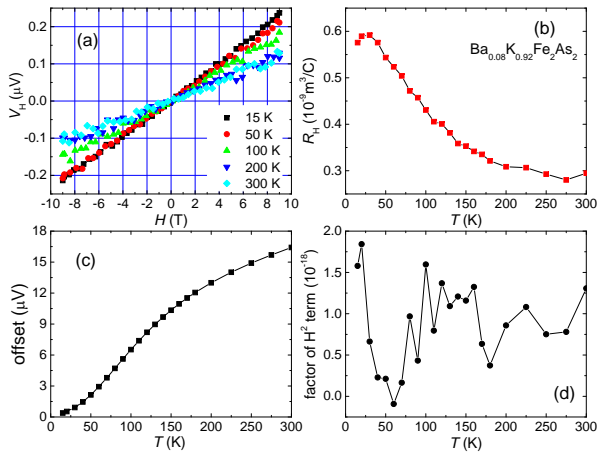}%
\caption{(Color online) (a) Hall voltage $V_H$ of Ba$_{1-x}$K$_x$Fe$_2$As$_2$ ($x$=0.92) single crystal measured by sweeping the field from -9 T to 9 T at selected temperatures. (b) Temperature dependence of Hall coefficient $R_H$ calculated from the linear term of the expression $V=V_{offset}+V_HH+V_{HH}H^{2}$. (c) The term $V_{offset}$ in the fit of raw data corresponds to the offset caused by longitudinal resistivity $\rho_{xx}$ between the Hall contacts. (d) The term $V_{HH}$ evaluates the contribution from $H^2$ term, which is very small and can be neglected. Solid lines are guides to the eye.}%
\label{illustration}%
\end{figure}

Let us turn to the normal-state transport properties. Figure \ref{illustration} illustrates an example how the analysis of Hall signal was processed for the sample $x$=0.92. The raw data can be decomposed into three terms as $V=V_{offset}+V_HH+V_{HH}H^{2}$, where $V_{offset}$ corresponds to the contribution of longitudinal resistivity $\rho_{xx}$ between the Hall contact, $V_H$ and $V_{HH}$ are hall voltages from the linear field dependent term and $H^2$ contribution, respectively. After subtracting the $V_{offset}$ term in the raw data, Figure \ref{illustration} (a) shows that the Hall voltage $V_H$ was measured as a function of applied field by sweeping the field from -9 T to 9 T at fixed temperatures. A nearly linear field dependence of $V_H$ is observed and the slopes $dV_H / dH$ retain positive values. The temperature dependence of Hall coefficient $R_H$ is shown in Fig. \ref{illustration}(b). The $V_{offset}$ term presents the temperature dependence of resistivity $\rho_{xx}$, as shown in Fig. \ref{illustration}(c). The good linear field dependence of raw data confirms the very weak contribution from $H^2$ term, as illustrated in Fig. \ref{illustration}(d).

\begin{figure}[tbh]%
\centering
\includegraphics[width=8cm]{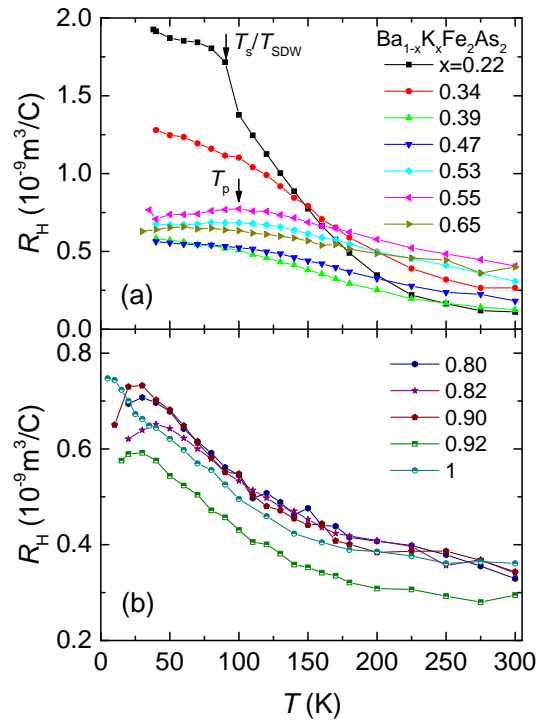}%
\caption{(Color online) Temperature dependence of Hall coefficient $R_H$ for Ba$_{1-x}$K$_x$Fe$_2$As$_2$ ($0.22 \leq x \leq 0.65$) (upper panel) and ($0.80 \leq x \leq 1$) (bottom panel) single crystals. Arrows indicate the kink corresponding to the SDW transition and the broad peak observed in the samples $x$=0.53, 0.55, and 0.65. Solid lines are guides to the eye.}%
\label{Hall coefficient}%
\end{figure}

Figure \ref{Hall coefficient} shows the temperature-dependent Hall coefficient $R_H$ of Ba$_{1-x}$K$_x$Fe$_2$As$_2$ ($0.22 \leq x \leq 1$) single crystals. As can be seen, for the underdoped sample $x$=0.22, $R_H$ shows a rapid increases with decreasing temperature, and becomes a plateau at $T$=100 K, where SDW transition occurs. For the sample $x$=0.34, $R_H$ gradually increases with decreasing temperature but shows a saturation tendency below $T$=150 K. With further increasing K doping levels, $R_H$ shows weak temperature dependence and a broad peak emerges at around 120-150 K for the samples $x$=0.47, 0.53, 0.55, and 0.65. All the samples $x$=0.34, 0.39, 0.47, 0.53, 0.55, and 0.65 show a convex temperature dependence below 200 K. As $x$ exceeds 0.80, the broad peak/big hump at around 120-150 K disappears and $R_H$ shows a rapid increases below $T$=150 K. A peak forms below $T$=50 K before the samples enter into superconducting state. $R_H$ follows a concave temperature dependence within the temperature range $50<T<300$ K.

The doping dependence of Hall effect reflects the change of relevant electronic structure \cite{Kittel,Ashcroft,Hurd}. The knowledge about band structure and its doping evolution in Ba$_{1-x}$K$_x$Fe$_2$As$_2$ system comes from ARPES measurements. Early ARPES data revealed that undoped ($x$=0) and optimally doped ($x$=0.4 and 0.45) samples have double-walled electron pocket at the $M$ points of BZ corner \cite{Ding,CLiu}. Zabolotnyy {\it et al.} \cite{Zabolotnyy} found that FS topology of BZ corner is actually characteristic of a propeller-shaped structure, which consists of five small FS sheets: a central circular pocket surrounded by four ‘‘blade’’ shaped pockets in Ba$_{1-x}$K$_x$Fe$_2$As$_2$ ($x$=0 and 0.3) single crystals. The central circular pocket around $M$ points is electronlike, while FS sheets around $\Gamma$ point and four blade pockets are holelike. The investigation on a wide doping range of Ba$_{1-x}$K$_x$Fe$_2$As$_2$ single crystals found that the gap size of the outer hole FS sheet around the BZ center shows an abrupt drop with overdoping (for $x \geq 0.6$) while the inner and middle FS gaps roughly scale with $T_c$ \cite{Malaeb}. In KFe$_2$As$_2$ single crystal, the FS around the BZ center was found to be qualitatively similar to that of Ba$_{0.6}$K$_{0.4}$Fe$_2$As$_2$ single crystal, but the electron pockets centered at $M$ points are completely absent due to an excess of hole doping \cite{Sato}. More detailed analysis of APRES data on the samples $x$=0.9 suggested the Lifshitz transition occurred between $0.7<x<0.9$ \cite{Xu}, which is supported by the theoretical calculations \cite{Khan}. Accordingly, the pairing symmetry was suggested to change from $s$ wave in optimal doped samples to $d$ wave in KFe$_2$As$_2$ \cite{Reid}. But most possibly, the superconducting gap structure changes from full gap state in the optimally doped samples into nodal-line structure state for KFe$_2$As$_2$ \cite{Okazaki,Hirano}.

It is noted that the broad peak/plateau in $R_H$ of Ba$_{1-x}$K$_x$Fe$_2$As$_2$ single crystals collapses in the overdoped samples ($0.80 \leq x \leq 1$), which coincides with the critical point where electron pocket disappears and Lifshitz transition occurs. The overall behavior of doping dependent $R_H$ is therefore related to the change of FS topology. Evtushinsky {\it et al.} \cite{Evtushinsky-Hall} had calculated the temperature dependence of Hall coefficient $R_H$ of optimally doped Ba$_{1-x}$K$_x$Fe$_2$As$_2$ based the propeller-like FS topology observed by ARPES experiments. The agreement suggested that the temperature dependence of Hall coefficient $R_H$ has the basis that FS evolves to propeller-like structure at low temperature regime. It should be pointed out that the same maximum of $R_H$ had been observed by Ohgushi {\it et al.} \cite{Ohgushi} in the Hall effect measurements on Ba$_{1-x}$K$_x$Fe$_2$As$_2$ ($0 \leq x \leq 0.55$) single crystals, which had been interpreted that an anomalous coherent state characterized by heavy quasiparticles in hole bands evolved below $T \sim 100$ K. The relative weak temperature dependence observed in the optimally doped samples may suggest that incoherence-coherence crossover is less pronounced. Our results strongly suggest that the maximum of $R_H$ observed within doping range $0.47 \leq x \leq 0.65$ as well as the temperature dependent behavior observed in the samples $x$=0.22, 0.34 and 0.39 are related to the contribution from the electron pocket at $M$ points of BZ. Without the contribution from the electron pocket, $R_H$ clearly drop at around $100<T<150$ K. In contrast to the electron doped Ba(Fe$_{1-x}$Co$_x$)$_2$As$_2$, where the hole contribution to the transport can be neglected at low temperatures in most of the phase diagram \cite{Rullier-Albenque}, electron conductivity plays a significant role in the charge transport of Ba$_{1-x}$K$_x$Fe$_2$As$_2$ below the doping $x$=0.80. The remarkable doping and temperature dependences of Hall coefficient $R_H$ in Ba$_{1-x}$K$_x$Fe$_2$As$_2$ system suggest a dominant interband interaction between carriers having electron and hole character \cite{Fanfarillo,Kemper}. 

We further analyze the Hall angle $cot \theta _H$=$\rho _{xx}$/$\rho _{xy}$ of Ba$_{1-x}$K$_x$Fe$_2$As$_2$ ($0.22 \leq x \leq 1$) single crystals. In our analysis, both longitudinal resistivity $\rho _{xx}$ and Hall resistivity $\rho _{xy}$ were normalized by their room temperature values. Therefore we have

\begin{equation}
cot \theta_{H} = \frac{\rho _{xx}}{\rho _{xy}} = \frac{\rho _{xx}}{R_HH} \propto \frac{\rho _{xx}/\rho _{xx} (300K)}{R_H / R_H (300K) }
\label{Hall angle equation}
\end{equation}

\begin{figure}[tbh]%
\centering
\includegraphics[width=8cm]{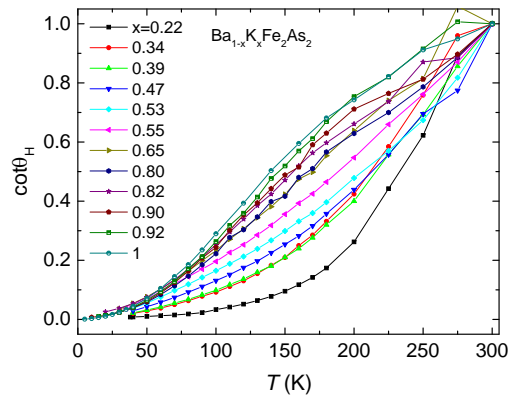}%
\caption{(Color online) Temperature dependence of Hall angle $cot \theta _{H}$ for Ba$_{1-x}$K$_x$Fe$_2$As$_2$ ($0.22 \leq x \leq 1$) single crystals. A concave temperature dependence is observed within the doping range $0.22 \leq x \leq 0.55$, whereas it dramatically changes to a convex temperature dependence within the doping range $0.65 \leq x \leq 1$. Solid lines are guides to the eye.}%
\label{Hall angle}%
\end{figure}

The detailed analysis of doping dependence of $\rho_{xx}$ can be found in Ref. \cite{Liu-BKFA}. The temperature dependence of Hall angle $cot \theta _{H}$ is shown in Fig. \ref{Hall angle}. Interestingly, the Hall angle data can be clearly divided into two groups. Hall angle $cot \theta _{H}$ displays a concave temperature dependence within the doping range $0.22 \leq x \leq 0.55$, whereas it changes to a convex temperature dependence within the doping range $0.65 \leq x \leq 1$. This feature again supports that the Lifshitz transition occurs at the critical doping $x=0.65 \sim 0.80$.

\begin{figure}[tbh]%
\centering
\includegraphics[width=8cm]{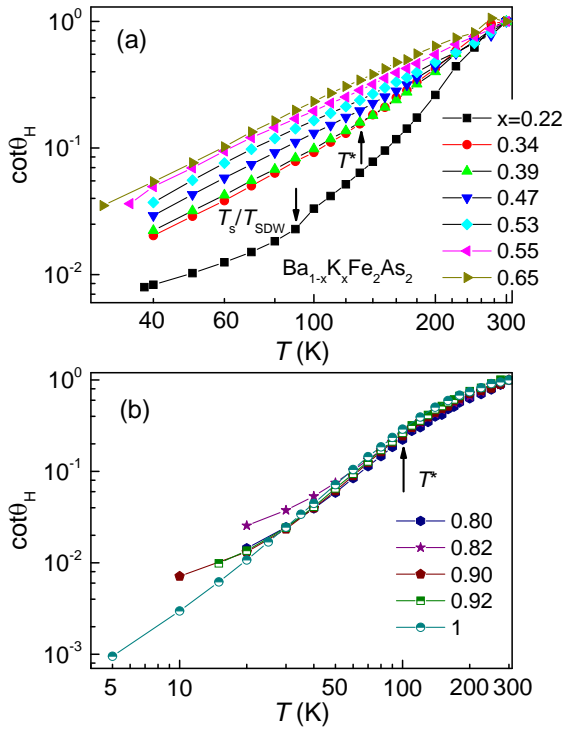}%
\caption{(Color online) Temperature dependence of Hall angle $cot \theta _{H}$ for Ba$_{1-x}$K$_x$Fe$_2$As$_2$ ($0.22 \leq x \leq 0.65$) (upper panel) and ($0.80 \leq x \leq 1$) (bottom panel) single crystals in double logarithmic plots. The arrows indicates the kink where slopes change for the samples $x$=0.34, 0.39, 0.47, 0.53, and 0.55 (a) and inflection point that $cot \theta _{H}$ has downward curvature above it and upward curvature below it for the samples $x$=0.80, 0.82, 0.90, 0.92, and 1 (b). Solid lines are guides to the eye.}%
\label{log plot Hall angle}%
\end{figure}

In an early work, the power-law temperature dependent Hall angle, i.e., $cot \theta_H$=$A+BT^{\alpha}$, was observed above a characteristic temperature $T$* in the entire phase diagram of Ba$_{1-x}$K$_x$Fe$_2$As$_2$ system \cite{YJYan}. Figure \ref{log plot Hall angle} shows our Hall angle data in double logarithmic plot. As can be seen, there is clear kink at around $T$*=140 K for the optimally doped samples $x$=0.34, 0.39, 0.47, and 0.53. For the different dopings, $T$* shifts little bit within the temperature range $120<T ^{\ast} <150$ K, which is quite close to the temperatures where $R_H$ and $d \rho _{xx}/dT$ \cite{Liu-BKFA} display the maximum. The slopes of the double logarithmic plots shown in Fig. \ref{log plot Hall angle} slightly change above and below $T$*. But we can see that temperature dependence of $cot \theta_H$ still follows the power law below $T$*. With doping approaching 0.65, the kink is smeared, and $cot \theta_H$ follows the power law within the whole temperature range above $T_c$. A different behavior is observed for the samples $x$=0.80, 0.82, 0.90, and 092. The power law (linear response) does not work well anymore. Above the characteristic temperature $T$*, $cot \theta _H$ displays the convex temperature dependence. But below $T$*, the concave temperature dependence is observed. Interestingly, we found that $cot \theta _H$ nearly follows $T^2$ dependence below $T$* for KFe$_2$As$_2$ single crystal. In fact, for high quality KFe$_2$As$_2$ single crystal, $\rho _{xx}$ follows a Fermi liquid behavior ($T^2$ dependence) below $T$=60 K, while $\rho _{xx}(300K)/ \rho _{xx}(4K)$ equals ~780 \cite{Liu-BKFA}. Meanwhile, $R_H$ only increases by a factor of 2 from 300 K to 5 K, i.e., $R_H$(5K)/$R_H$(300K) $\sim$ 2. Therefore, longitudinal resistivity $\rho _{xx}$ actually dominates the behavior of  $cot \theta_H$, which leads to $T^2$ dependence of  $cot \theta_H$ at low temperature regime.

Finally, we discuss the correlation among magnetic susceptibility, Hall coefficient, and resistivity in Ba$_{1-x}$K$_x$Fe$_2$As$_2$ compounds. Recently, Nakajima {\it et al.} reported the study of normal-state charge dynamics in BaFe$_2$(As$_{1-x}$P$_x$)$_2$, Ba(Fe$_{1-x}$Co$_x$)$_2$As$_2$, and Ba$_{1-x}$K$_x$Fe$_2$As$_2$ through the measurements of optical conductivity spectrum and resistivity \cite{Nakajima}. For BaFe$_2$As$_2$, charge dynamics is incoherent at $T$=300 K. The decomposition of the optical conductivity spectrum of KFe$_2$As$_2$ is nearly the same as that of BaFe$_2$As$_2$. The highly incoherent spectrum seems to persist over the entire doping range in the normal state of Ba$_{1-x}$K$_x$Fe$_2$As$_2$ system. The results strongly suggest that quasiparticle states on a substantial part of FS remain incoherent at high temperatures in Ba$_{1-x}$K$_x$Fe$_2$As$_2$ system \cite{Nakajima}. Taking the two-fluid model suggested for magnetic response of heavy electron materials \cite{Yang,Shirer}, the local moment spins dominate above the peak temperature $T$*, whereas the hybridization of local moment spins and conduction electron spins is significant and contributes more to magnetic susceptibility below $T$*. Coherent component plays a significant role below $T$*, where both resistivity and susceptibility drop \cite{Nakajima,Haule09}]. In fact, the temperature dependence of the magnetic susceptibility and the thermal expansion provide experimental evidence for the existence of a coherence-incoherence crossover in KFe$_2$As$_2$ \cite{Hardy}. The broad maximum at around 120 K indicates the onset of coherence. In the optimal doping region $0.34 \leq x \leq 0.47$, SDW order is suppressed while monotonic increase of magnetic susceptibility extends to 800 K. The broad hump emerges at $x$=0.53 and evolves into a broad peak at around 120 K in KFe$_2$As$_2$. Our magnetic susceptibility data suggest that superconductivity with high transition temperature emerges when the incoherence-coherence crossover is less pronounced in Ba$_{1-x}$K$_x$Fe$_2$As$_2$ system.

Resistivity of Ba$_{1-x}$K$_x$Fe$_2$As$_2$ superconductors shows a tendency for saturation above 100 K, which gives rise to a broad peak in the plots of $d\rho_{ab} /dT$ vs $T$ \cite{Liu-BKFA}. This characteristic temperature is in coincidence with the peak temperature of susceptibility curves. Hall coefficient, $R_H$ displays weak doping and temperature dependences above 150 K. But low temperature part within the doping range $0.80 \leq x \leq 1$ is quite distinct from that of the samples $0.22 \leq x \leq 0.65$. $R_H$ tends to saturate below 150 K for the samples $0.22 \leq x \leq 0.65$, whereas it shows rapid increase for the samples $0.80 \leq x \leq 1$. It should be emphasized that the analysis of Hall angle also supports the existence of characteristic temperature $T$*, which is suggested to be related to the incoherence-coherence crossover. Assuming two types of charge careers in Ba$_{1-x}$K$_x$Fe$_2$As$_2$ system \cite{Nakajima}, above $T$*, highly incoherent charge carriers dominates, whereas coherent ones become significant below it. Here the coherence process is related to the hybridization of conduction charge carriers and local spin moments, which gives rise to a large effective mass of conduction charge carriers. The overall behavior of magnetic susceptibility, Hall coefficient, and resistivity provides evidences of incoherence-coherence crossover at $T$* in Ba$_{1-x}$K$_x$Fe$_2$As$_2$ system. The coherent charge dynamics in BaFe$_2$(As$_{1-x}$P$_x$)$_2$ and Ba(Fe$_{1-x}$Co$_x$)$_2$As$_2$ systems is more pronounced than Ba$_{1-x}$K$_x$Fe$_2$As$_2$ system in the normal state \cite{Nakajima}. It could be the reason why the coherence-incoherence crossover is not observed in resistivity and magnetic susceptibility of BaFe$_2$(As$_{1-x}$P$_x$)$_2$ and Ba(Fe$_{1-x}$Co$_x$)$_2$As$_2$ systems.

\section{Conclusions}

In summary, we have performed magnetic susceptibility $\chi (T)$ and Hall coefficient $R_H$ measurements on a series of Ba$_{1-x}$K$_x$Fe$_2$As$_2$ single crystals. A crossover from SDW ordered sate to KFe$_2$As$_2$ -type magnetic interactions occurs with increasing K content. It is found that $\chi (T)$ monotonically increases with increasing temperature for the underdoped and optimally doped samples $0.13 \leq x \leq 0.47$. For the overdoped samples $0.53 \leq x \leq 1$, a big hump was observed at around 150 K, and it eventually evolves into a broad peak in KFe$_2$As$_2$ at 120 K. The magnitude of magnetic susceptibility keeps increasing with increasing K content. Hall coefficient $R_H$ and Hall angle $cot \theta_H$ display the dramatic change as $x$ exceeds 0.80, which coincides with the critical doping point where electron pocket disappears with excess hole doping. Our results strongly support that the change of doping dependence of Hall coefficient $R_H$ and Hall angle $cot \theta_H$ is related to the change of FS topology, i.e. the Lifshitz transition. The characteristic temperature $T$* is identified in magnetic susceptibility, Hall coefficient, and resistivity data, which strongly suggests the incoherence-coherence crossover occurred in Ba$_{1-x}$K$_x$Fe$_2$As$_2$ system.

\section{Acknowledgements}

This work was supported by the U.S. Department of Energy (DOE), Office of Science, Basic Energy Sciences, Materials Science and Engineering Division. The research was performed at the Ames Laboratory, which is operated for the U.S. DOE by Iowa State University under contract DE-AC02-07CH11358.


\end{document}